# High-resolution reflection spectra for Proxima b and Trappist-1e models for ELT observations

**Authors:** Lin, Z.[1,2]*, L. Kaltenegger[1,2]

**Affiliations:**

[1]Carl Sagan Institute, Cornell University, Ithaca, NY 14853, USA.

[2]Astronomy Department, Cornell University, Ithaca, NY 14853, USA.

*Correspondence to: zl433@cornell.edu

**Abstract:** The closest stars that harbor potentially habitable planets are cool M-stars. Upcoming ground- and space-based telescopes will be able to search the atmosphere of such planets for a range of chemicals. To facilitate this search and to inform upcoming observations, we model the high-resolution reflection spectra of two of the closest potentially habitable exoplanets for a range of terrestrial atmospheres and surface pressures for active and inactive phases of their host stars for both oxic and anoxic conditions: Proxima b, the closest potentially habitable exoplanet, and Trappist-1e, one of 3 Earth-size planets orbiting in the Habitable Zone of Trappist-1. We find that atmospheric spectral features, including biosignatures like $O_2$ in combination with a reduced gas like $CH_4$ for oxic atmospheres, as well as climate indicators like $CO_2$ and $H_2O$ for all atmospheres, show absorption features in the spectra of Proxima b and Trappist-1e models. However for some features like oxygen, high-resolution observations will be critical to identify them in a planet's reflected flux. Thus these two planets will be among the best targets for upcoming observations of potential Earth-like planets in reflected light with planned Extremely Large Telescopes.

## 1. Introduction

Over 4,000 exoplanets have been found to date. Among these planets, some are located in the circumstellar habitable zone (HZ), which should allow surface liquid water to exist (see e.g. Udry et al. 2007; Borucki et al. 2011, Batalha et al. 2013; Kaltenegger et al. 2013; Quintana et al. 2014; Kane et al. 2016; Kaltenegger 2017). In the search for potentially habitable worlds, M dwarfs are of particular interest, because they are the most abundant types of stars in our galaxy, and make up about 75% of stars in the solar neighborhood. Earth-like planets orbiting small M dwarfs have both shorter transit periods and deeper transit signals than Earth-like planets orbiting solar analog stars. Furthermore, the frequency of rocky planets orbiting cool stars appears to be higher, placing such planets orbiting close-by stars among the most suitable targets for follow-up studies of their atmosphere in the near future (see e.g. Scalo et al 2007; Dressing & Charbonneau 2015).

The closest detected potentially habitable worlds orbit such red stars. Here we present high-resolution reflection spectra in the visible to near-infrared (0.4 to 5.0 μm) wavelength ranges for Proxima b and Trappist-1e (see Table 1), based on a range of atmosphere models described in O'Malley-James & Kaltenegger (2019a): from oxic atmosphere with varying surface pressures, including both Earth-like and eroded atmospheres, to anoxic atmospheres, which mimic an early Earth atmosphere before the rise of oxygen. We also discuss how different planetary surface would influence the planets' flux.

Several studies have addressed the habitability of Proxima b and Trappist-1e in terms of water inventories, atmospheric mass and composition (e.g. Lincowski et al 2018, Wolf 2017, Ribas et al 2016, Barnes et al 2016,

Goldblatt et al. 2017, Turbet et al 2016, Dong et al 2017, Ramirez & Kaltenegger 2018).

The surface pressure of rocky exoplanets is unknown. In addition the extent of atmospheric erosion Proxima b and Trappist-1e have experienced is difficult to quantify without information on the planet's magnetic field, the stellar wind pressure at the planet's orbit, and the atmospheric composition. Therefore we included models for different surface pressure, accounting for atmosphere erosion and different initial surface pressures in O'Malley-James & Kaltenegger (2019a), which is the base of our spectra shown here.

While M-stars can be active (e.g. West et al. 2011) and higher amounts of UV can hit their planets, several teams have made the case that planets in the HZ of M stars can remain habitable, despite periodic high UV fluxes (see e.g. discussion in O'Malley-James & Kaltenegger 2019a, Rimmer et al. 2018, Segura et al. 2010, Scalo et al. 2007, Buccino et al. 2007, Tarter et al. 2007, Heath et al. 1999). Note that recent studies suggest that high UV surface flux may even be necessary for prebiotic chemistry to occur (see Ranjan & Sasselov 2016; Rimmer et al. 2018).

Observations of Proxima b and Trappist-1e are planned with several upcoming and proposed telescopes (e.g. Batalha et al 2018, Rodler & Moralez 2014, Snellen et al. 2013) both on the ground and in space like the James Webb Space Telescope (JWST) and the Extremely Large telescopes (ELTs), such as the Giant Magellan Telescope (GMT), Thirty Meter Telescope (TMT), and the Extremely Large Telescope (ELT) and several missions concepts like Origins (Battersby et al. 2018), Habex (Mennesson et al. 2016) and LUVOIR (LUVOIR Team, 2018). Future ground-based ELTs and JWST are designed to obtain the first measurements of the atmospheric composition of Earth-sized planets (see e.g. Kaltenegger & Traub 2009; Kaltenegger et al. 2011, Hedelt et al. 2013; Snellen et al. 2013; Rodler & Lopez-Morales 2014, Stevenson et al. 2016; Barstow & Irwin 2016).

Here we use the 39m diameter ELT as an example for near-future ground-based telescopes. With its expected inner working angle of 6 milliarcsecond (mas) for visible wavelengths, it will be able to resolve Proxima b, with an apparent angular separation of 37 mas from its host star.

Note that Trappist-1e, with an apparent angular separation of 2.4 mas, will not be resolved by ELT. However, ground-based high-resolution spectroscopy has already characterized several atmospheric species like carbon monoxide and water vapor in the atmosphere of unresolved planets like HD 179949 b (see e.g. Brogi et al. 2014), which was characterized at the Very Large Telescope (VLT) by using the planet's known changing Doppler shift during the observations.

Proxima b and Trappist-1e are both intriguing targets for observations to characterize their atmospheres. Because of Proxima b's large apparent angular separation, it can be resolved by planned ground-based telescopes. Therefore, its status as a resolved planet around the closest star to the Sun makes it one of the best targets to study in the near future. This paper is structured as follows: section 2 discusses our model, section 3 presents the spectra and contrast ratios, section 4 discusses our results and section 5 is our conclusion.

## 2. Methods

Our stellar input spectra at the location of Proxima Centauri and Trappist-1, shown in Fig.1 and Fig. 2 (top row), are based on PHOENIX models (Husser et al. 2013) in combination with IUE data[1] (see O'Malley-James & Kaltenegger 2019a for details). We model the high-resolution reflection spectra for a range of atmospheres for each of the planets: (i) a 1 bar surface pressure atmosphere assuming Earth-like mixing ratio; (ii) eroded atmospheres with 0.5 bar and 0.1 bar surface pressures assuming Earth-like mixing ratio, and (iii) an anoxic atmosphere (trace levels of $O_2$; $3 \times 10^{-3}$ $CO_2$) with 1 bar surface pressure that mimics Earth's atmosphere before the Great Oxidation Event (see details in O'Malley-James & Kaltenegger 2019a). The key assumptions for these models are summarized in Table 2. To maintain surface temperatures above freezing for the 1 bar oxic cases of Proxima b, we assume a constant mixing ratio of 100 times present atmospheric levels of $CO_2$ ($3.65 \times 10^{-2}$) for both planets and a $CH_4$ mixing ratio of $1.6 \times 10^{-6}$ at the surface of Trappist-1e (M8.0V host), and $1.6 \times 10^{-4}$ at the surface of Proxima b (M5.5V host), which receives slightly less incident radiation due to its orbital separation.

---

[1] http://archive.stsci.edu/iue/

We use Exo-Prime, a coupled one-dimensional model developed for rocky exoplanets (see details in Kaltenegger & Sasselov 2010). The line by line radiative transfer model, which generates the spectra, is based on a model originally developed to observe trace gases in the stratosphere of Earth (Traub & Stier 1976, Jucks et al. 1998), and has been further developed to model exoplanet spectra (see e.g. Kaltenegger et al. 2007, 2013; Kaltenegger & Traub 2009). We divide the exoplanet atmospheres into 35 layers for our models up to an altitude of at least 60 km, with smaller spacing towards the ground. The atmospheric species, which account for the most significant spectral features are $H_2O$, $CO_2$, $O_2$, $H_2$, $CH_4$, $CO$, $N_2O$, $CH_3CL$, $OH$, $O_3$ for the oxic cases, and $H_2O$, $CO_2$, $O$, $O_2$, $H$, $OH$, $HO_2$, $H_2O_2$, $O_3$, $H_2$, $CO$, $HCO$, $H_2CO$, $CH_4$, $CH_3$, $C_2H_6$, $NO$, $NO_2$, $HNO$, $SO$, $SO_2$, $H_2SO_4$ for the anoxic cases. In our models we assume an Earth-like surface albedo with 70% ocean, 2% coast, and 28% land with 50% cloud coverage. The land surface is divided into 30% grass, 30% trees, 9% granite, 9% basalt, 15% snow, and 7% sand (following Kaltenegger, Traub & Jucks 2007). In the discussion section we also discuss the influence of different surfaces in the reflected spectra of these planets.

Clouds generally increase the reflectivity while obscuring surface and deeper atmosphere layers, thus clouds can have a strong impact on detectability of atmospheric species. Note that the properties and height of cloud layers require knowledge of unknown attributes like topography and rotation rate. While global climate models have started to expand to model these effects, so far the results are not conclusive, therefore we use an Earth-like cloud structure only to explore the effect of clouds on the spectra.

We model the reflection spectra at high-resolution, with a step size of 0.01 $cm^{-1}$ wavenumber from 0.4 to 5 μm providing a minimum resolving power of $\lambda/\Delta\lambda = 100,000$ at all wavelengths, which corresponds to the proposed spectral resolution of the High-resolution Spectrograph (HIRES) built for the ELT. From about 4 μm onwards, the thermal emission of the two planets becomes comparable to the reflected planetray flux. While we include the emission spectra of the planets in our models, due to low overall flux in these wavelengths, the thermal emission does not add new detectable spectral feature of interest to the spectra. For clarity we present the spectra in this paper at a resolution of $\lambda/\Delta\lambda = 300$, smeared using a triangular smoothing kernel. High-resolution spectra for all models are available online (http://carlsaganinstitute.org/data/).

## 3. Results

Temperature and chemical mixing ratio profiles for our models are described and shown in O'Malley-James & Kaltenegger (2019a). We summarize the major model characteristics of that paper here to link them to the atmospheric features that are shown in the spectra and contrast ratios: for the three oxic atmosphere models, temperature increases as surface pressure increases. $CO_2$ is well mixed in the atmospheric models and is set to a mixing ratio of $3.0 \times 10^{-3}$ for the anoxic models and $3.65 \times 10^{-2}$ for the oxic models. The $CH_4$ mixing ratio on the surface is $3.0 \times 10^{-7}$ for the anoxic models and $1.6 \times 10^{-6}$ for the Trappist-1e, and $1.6 \times 10^{-4}$ for the Proxima b oxic models, respectively. For the oxic atmosphere models of both planets, the $H_2O$ mixing ratio is the highest for the 0.1 bar case, due to increased evaporation as surface pressure decreases. For the anoxic 1bar atmosphere models, the $H_2O$ mixing ratio is higher than for the 1 bar oxic model due to warmer surface temperature and consequential increased evaporation. The mixing ratio of $O_3$ slightly increases for the planetary models for active stars (O'Malley-James & Kaltenegger 2019a), due to increased UV irradiation, while the concentration of $CH_4$ shows small variation for an active versus inactive stellar irradiation for both planets. For the anoxic atmosphere models, $CH_4$ concentration decreases by orders of magnitude in the upper atmosphere of both planets for active host stars input spectra.

### 3.1 Oxic atmosphere models

Fig. 1 shows the reflection spectra and the contrast ratio for Earth-like atmospheres from 1 bar surface pressure to eroded atmosphere models of 0.1bar for Proxima b. Fig. 2 shows the reflection spectra assuming similar atmosphere models for Trappist-1e. Fig. 1 and Fig. 2 (top row) shows the stellar flux of the host stars compared to the Sun's (dashed line). The two middle panels show the reflection spectra of the planet models for active stellar spectra as (middle top) absolute show flux (middle top) and contrast ratio of the planet to its host star's flux ratio (middle bottom).

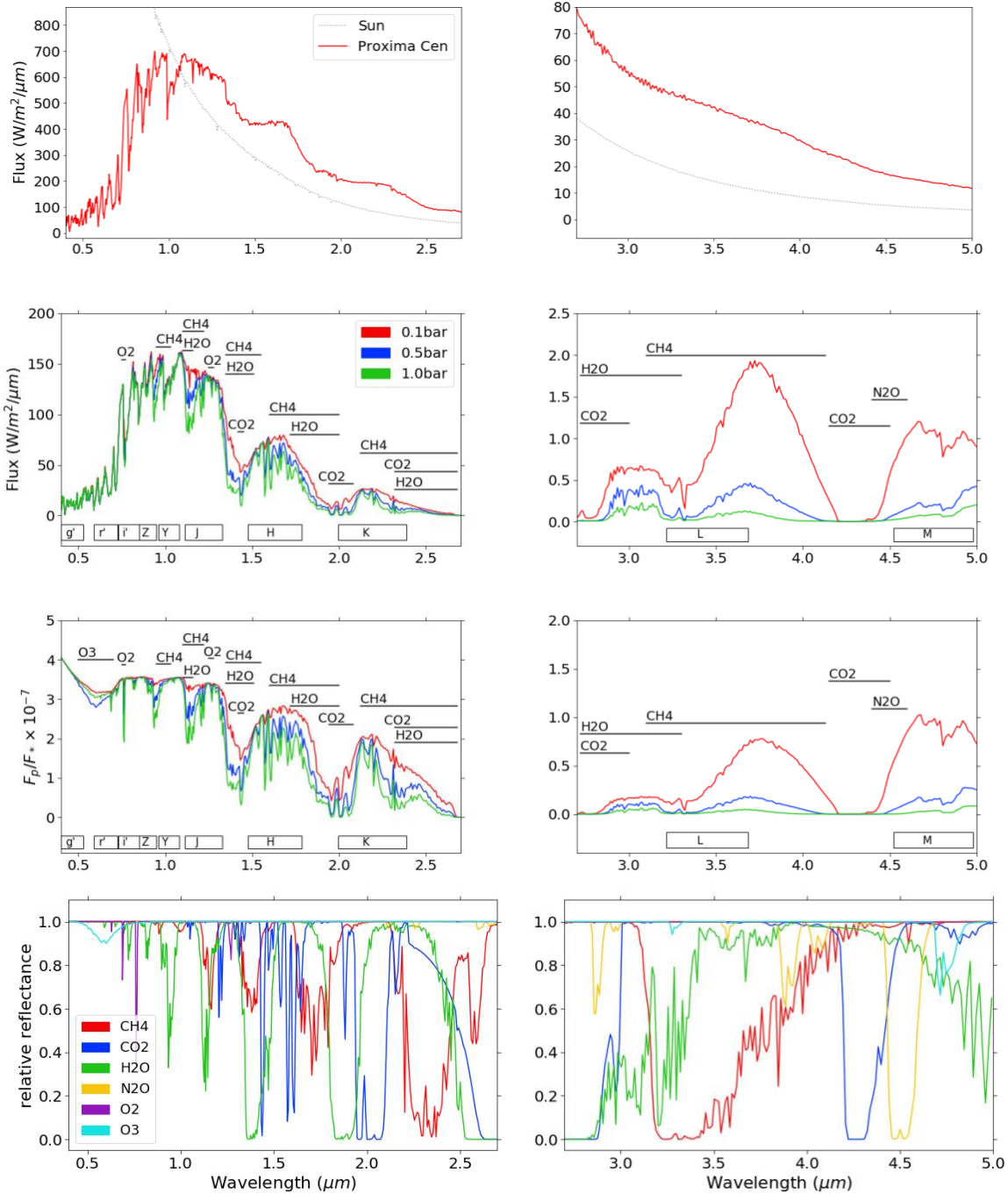

**Figure 1:** Reflection spectra for Proxima b (lower three rows) and stellar spectrum of its host star Proxima Centauri (top). Reflection spectra (middle top) and planet to star contrast ratio (middle bottom) show three oxic planet models: an Earth-like 1 bar (green), an eroded 0.5 bar (blue), and an eroded 0.1 bar (red) surface pressure atmosphere. The bottom row shows the absorption by individual molecules for the 1bar surface pressure model to highlight position and overlap of absorption features. Note that spectra from 0.4 to 2.7 µm (left) and spectra from 2.7 to 5.0 µm (right) are shown in different scale for clarity. All spectra are smeared to a resolving power of 300.

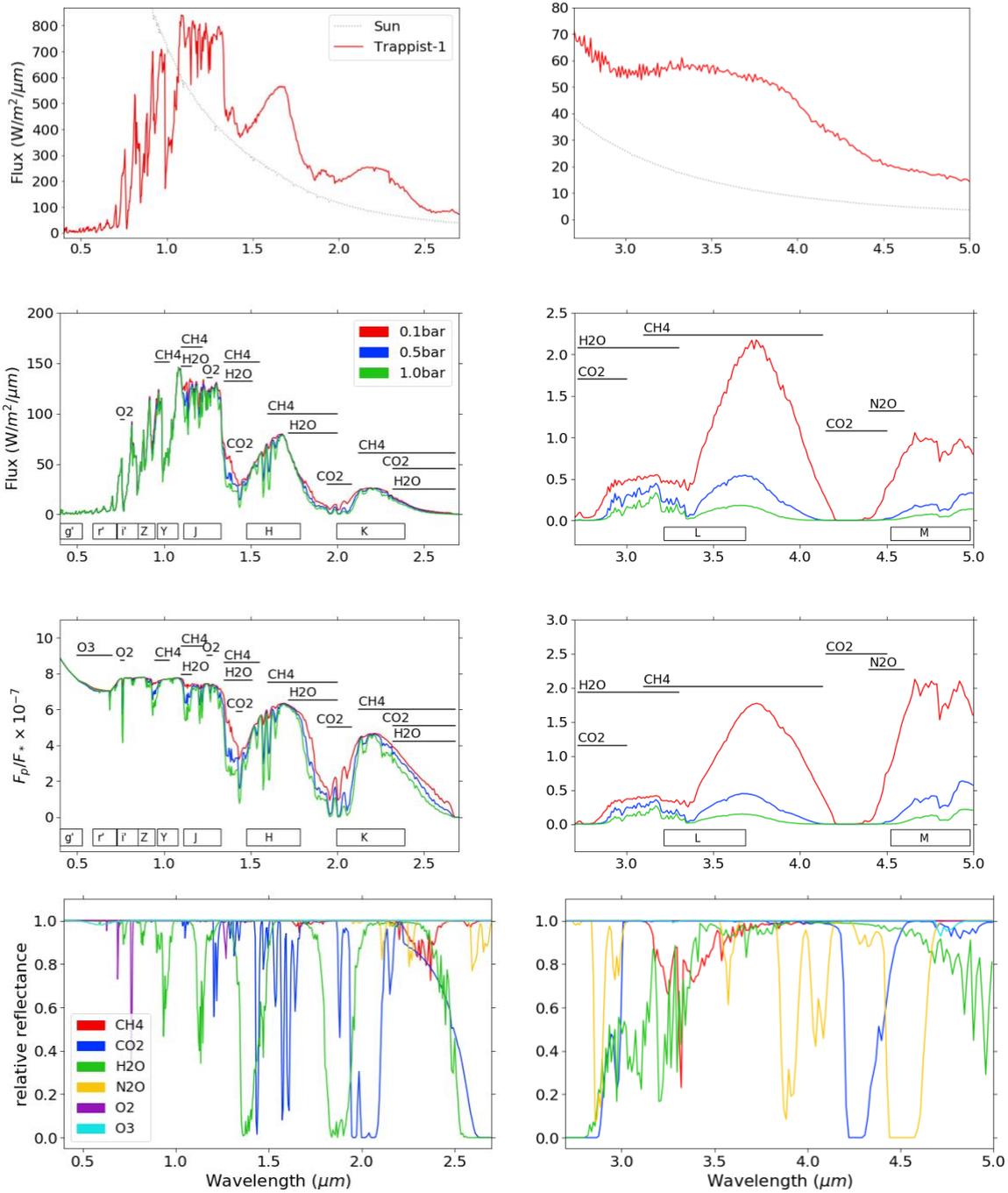

**Figure 2: Reflection spectra for Trappist-1e (lower three rows) and stellar spectrum of its host star Trappist-1 (top). Reflection spectra (middle top) and planet to star contrast ratio (middle bottom) show three oxic planet models: an Earth-like 1 bar (green), an eroded 0.5 bar (blue), and an eroded 0.1 bar (red) surface pressure atmosphere. The bottom row shows the absorption by individual molecules for the 1bar surface pressure model to highlight position and overlap of absorption features. Note that spectra from 0.4 to 2.7 μm (left) and spectra from 2.7 to 5.0 μm (right) are shown in different vertical scale for clarity. All spectra are smeared to a resolving power of 300.**

Each reflection spectra panel contains three atmospheric models, corresponding to a 1.0 bar (green), an eroded 0.5 bar (blue), and an eroded 0.1 bar surface pressure model (red). The bottom panel of Fig. 1 and Fig. 2 show the relative absorption of the individual chemicals in the planet's 1bar surface pressure atmosphere to clarify overlap of individual spectral lines. These individual chemical plots are generated by removing the opacity of all other molecules from the original atmosphere, and calculating only the opacity due to one particular atmospheric species (see Kaltenegger & Traub 2007). Note that we split the wavelengths into two ranges, one from 0.4 to 2.7 μm, and one from 2.7 to 5.0 μm, because in longer wavelengths the spectra have much lower flux, and therefore we split the model spectra of both planets into two wavelength ranges to show the absorption features clearly, which requires different scales.

A general trend shown in Fig. 1 and Fig. 2 is that as pressure increases, the depth of the absorption features increases, which is expected because for reflection spectra the absolute abundance of chemicals in the atmosphere determines the depth of the spectral absorption features. Fig. 1 and Fig. 2 shows that for both model planets, most notably the water absorption features decrease with decreasing pressure due to the lower absolute amount of $H_2O$ in the planet model atmospheres. Absorption features at 1.1, 1.4, 1.8, and 2.7 μm for $H_2O$, at 1.6, 2.0, and 2.7 μm for $CO_2$, at 1.7, 2.4, and 3.3 μm for $CH_4$, and at 4.5 μm for $N_2O$ can be seen in all planet reflection spectra.

The incident stellar flux determines the absolute reflected flux from the planet and thus the depth of absorption features, which is clearly illustrated in the comparison of the reflected spectra and contrast ratio in the middle panels of Fig. 1 and Fig. 2. The contrast ratio plot shows a wider variety of chemical absorption including $O_2$ at 0.76 μm and $O_3$ at 0.45 - 0.74 μm because the depths of the absorption features in reflected flux shown as contrast ratio between the planet and its host star are not modulated by the incident stellar flux.

Note that several of the $CO_2$, $CH_4$ and $H_2O$ and $N_2O$ features overlap for a resolution of 300 (bottom row, Fig. 1 and Fig. 2). It will require either a wider wavelength coverage to identify them in non-overlapping bands or a higher spectral resolution to tell them apart.

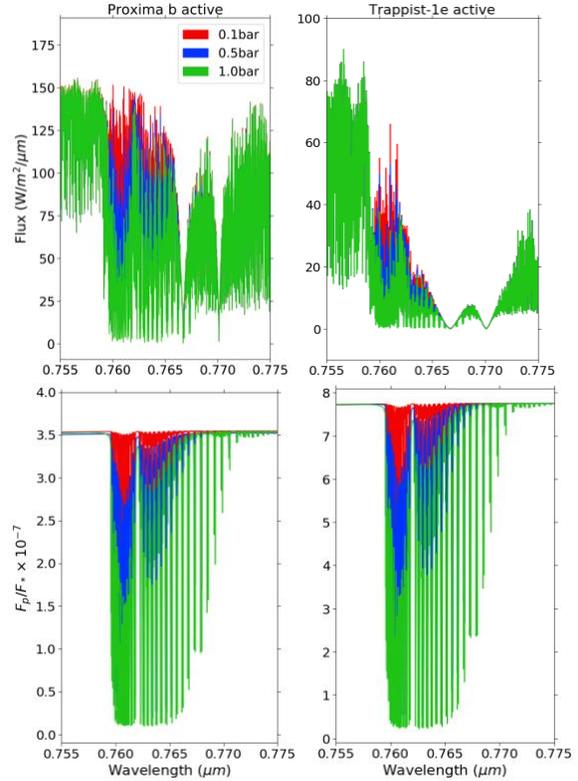

**Figure 3:** High-resolution reflection spectra of Proxima b and Trappist-1e focused on the $O_2$ feature at 0.76 μm for Earth-like oxic atmosphere models with three surface pressures of 1.0 bar (green), 0.5 bar (blue), and 0.1 bar (red) shown as absolute planetary flux (top) and as planet-star contrast ratio (bottom).

Absorption features for $O_2$ (or $O_3$) in combination with $CH_4$ (see e.g. Lovelock et al. 1965; Lederberg 1965; Lippincott et al. 1967) indicates life on Earth and is used as a spectral biosignature. It can be seen in Fig. 1 and Fig. 2 for the oxic models of both planets in the contrast ratio plots of planetary versus stellar flux, but not in the reflection spectra shown at a resolution of 300 because the incident flux levels of both stars are very low at these wavelengths. In high-resolution of 100,000 (Fig. 3) the $O_2$ feature at 0.76 μm for all modelled surface pressure levels can be seen in the oxic atmosphere models even in reflected planetary flux (Fig. 3 top) and clearly in the contrast ratio plot (Fig. 3 bottom).

The reflection spectra for oxic models of both planets for active versus inactive host stars show very little variations and are therefore not shown separately here but can be downloaded from the high-resolution database of spectra for

Proxima b and Trappist-1e (http://carlsaganinstitute.org/data/). Note that $O_3$ concentrations show a slight stellar activity dependence in our oxic models for both Proxima Centauri and Trappist-1, increasing with stellar activity as expected, which in turn slightly increases the depth of Chappuis band from 0.45-0.74 μm (see also Rugheimer & Kaltenegger 2018).

### 3.2 Anoxic atmosphere models

The reflection spectra for anoxic 1 bar atmospheres for Proxima b (Fig. 4 left column) and Trappist-1e (Fig. 4 right column) are dominated by absorption features of $H_2O$ at 1.4, 1.9, 2.7, 1.1, and 0.9 μm (sorted here by decreasing absorption feature strength in reflection spectra) and $CO_2$ at 2.0, 2.7, 1.6, and 1.4 μm. Fig. 4 shows that $CH_4$ only contributes slightly to the reflection spectra for anoxic atmospheres (bottom, $CH_4$ indicated in red). The bottom panel of Fig. 4 shows the absorption by individual molecules to highlight position and overlap of absorption features. In the low resolution of 300, the 1.4 μm and 2.0 μm $CO_2$ feature overlap with $H_2O$ features, which will require either a wider wavelength coverage to identify them in non-overlapping bands or high spectral resolution to tell them apart. Anoxic spectra from 2.7 to 5 μm are shown in Fig. 5.

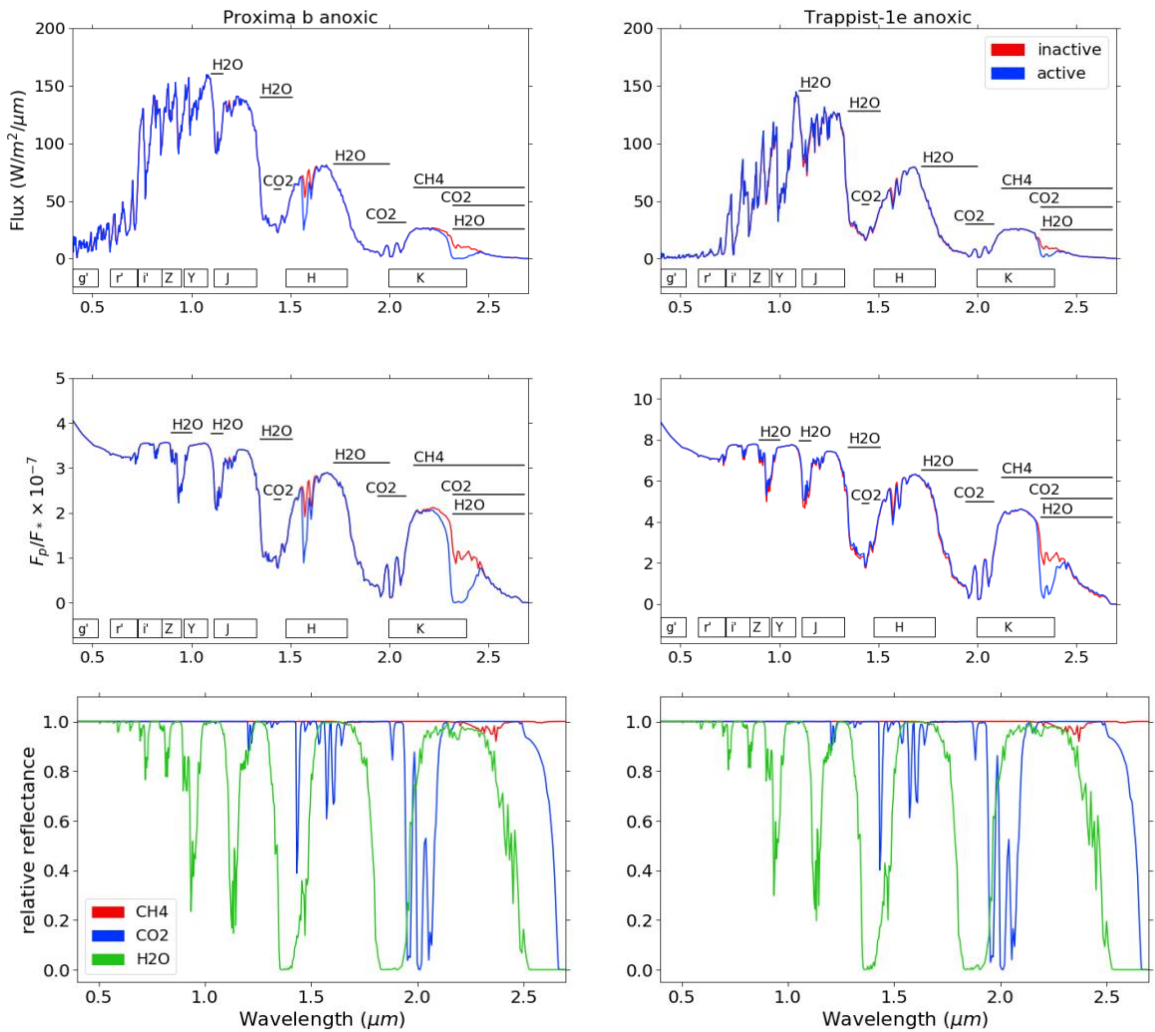

**Figure 4: Reflection spectra for Proxima b (left column) and Trappist-1e (right column), assuming an anoxic atmosphere show for inactive (red), and active (blue) host star spectra. The bottom row shows the absorption by individual molecules. All spectra are smeared to a resolving power of 300.**

We compare the reflection spectra of anoxic atmosphere models of Proxima b to Trappist-1e in Fig. 5, to show how stellar spectra and relative size of planet and host star can influence the spectra and contrast ratio. For the reflected planetary flux (Fig. 5 top), Proxima b and Trappist-1e have similar flux magnitude, because their irradiance only differs by about 1.2% times Earth's irradiance (Table 1). From 0.4 to 1 μm, Proxima b has higher reflected planetary flux, due to higher incident stellar irradiation at these wavelengths from its hotter host star. The contrast ratio of Trappist-1e is about twice that of Proxima b due to its larger relative size compared to its host star. However, because of its larger apparent angular separation, the contrast ratio of Proxima b can be enhanced by a factor of $10^3$-$10^4$ (Lovis et al. 2017) compared to an unresolved planet like Trappist-1e, making Proxima b an intriguing target for observation in the near future.

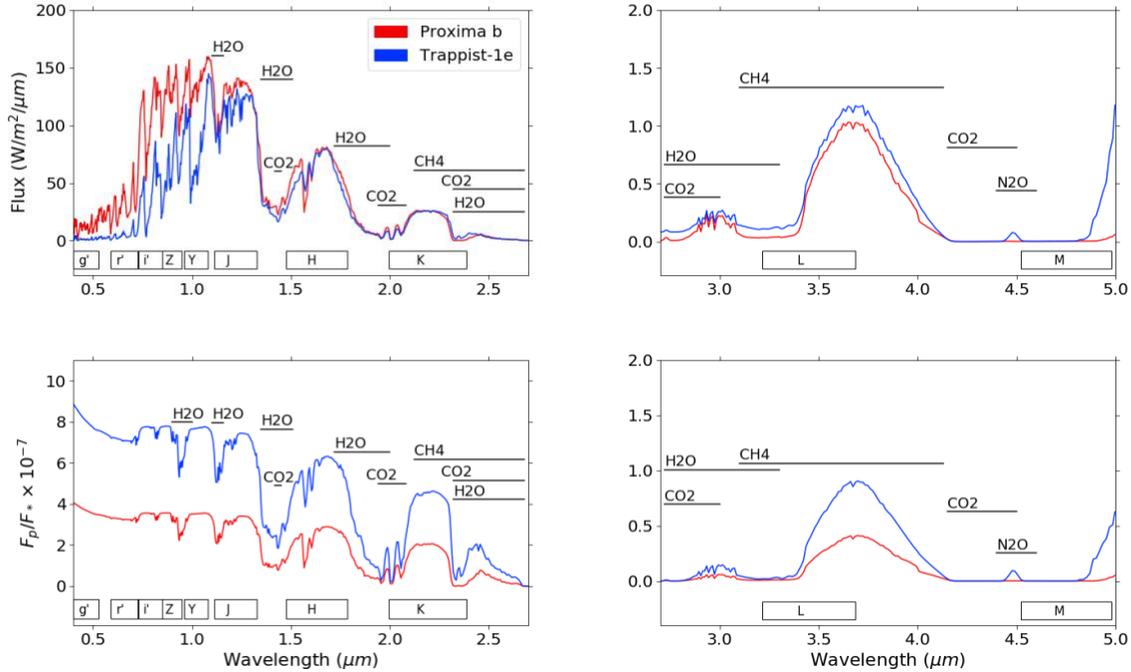

**Figure 5: Spectra of anoxic atmosphere models of Proxima b (red) and Trappist-1e (blue). Note that spectra from 0.4 to 2.7 μm (left) and spectra from 2.7 to 5 μm (right) are shown in different vertical scales for clarity. All spectra are smeared to a resolving power of 300.**

## 4. Discussion

### 4.1 Activity is an open question regarding surface Habitability of Proxima b and Trappist-1e

Surface habitability of planets orbiting M dwarfs has been questioned because of the intensity and frequency of UV activity of their host stars, especially for young M stars (see e.g. discussion in Scalo et al. 2007, Tarter et al. 2007, Shields et al. 2016, France et al. 2016, Kaltenegger 2017, Loyd et al. 2018, Günther et al. 2019). Besides, low mass M stars can remain active for an extended period of time.

Stellar UV activity can increase a HZ planet's surface UV flux by up to two orders of magnitude (see e.g. Segura et al. 2010; Tilley et al. 2019). Frequent strong flares of Trappist-1 have been raised as a concern for its suitability for life (e.g. Vida et al. 2017, Yamashiki et al. 2019). High energy superflares can also occur on an annual basis on Proxima Centauri (e.g. Vida et al. 2019), which could affect the surface habitability of Proxima b.

High UV radiation can damage biological molecules including nucleic acids on a planet's surface (see e.g. Kerwin & Remmele 2007). In addition, high energy particle fluxes produced by flaring events could cause erosion of a planet's atmosphere as well as water loss over time (e.g. Vidotto et al. 2013; Garraffo et al. 2016; Kreidberg & Loeb 2016; Ribas et al. 2016; Turbet et al. 2016;

Airapetian et al. 2017; Dong et al. 2017; Garcia-Sage et al. 2017; Kopparapu et al. 2017; Lingam & Loeb 2017; Barnes et al. 2018; Goldblatt 2018; Meadows et al. 2018; Lammer et al. 2007; See et al.2014).

On surface of planets orbiting M stars, high UV environments could remain for billions of years (see e.g. West et al 2004, France et al 2013, Rugheimer et al 2015, Youngblood et al 2016). Therefore, protective mechanisms that allow organisms to survive such environment would be essential for maintaining surface habitability. For the eroded and anoxic atmospheres, due to low optical depth of UV shielding gas, such mechanisms are of greater importance.

Studies of extremophiles on Earth identified several strategies organisms apply to survive high energy radiation. Protective pigments, for example, could attenuate incoming radiation, and DNA repair pathways can reduce or even prevent damages due to radiation (see e.g. Neale & Thomas 2016; Sancho et al 2007; Onofri et al., 2012; Cockell et al., 1998). Living subsurface, such as under a layer of rock, soil, sand, or water, can significantly reduce the exposure to detrimental UV radiation, and therefore increase habitability (e.g. Ranjan & Sasselov 2016; Cockell et al 2000,2009, O'Malley-James & Kaltenegger 2017). However, this strategy would make remote detection of such life difficult. Biofluorescence is an alternative protective mechanism. Widely observed in nature, biofluorescence can also convert UV light into less energetic wavelengths, therefore protecting the organisms from damage and increasing the detectability of such a biosphere (O'Malley-James & Kaltenegger 2018, 2019b).

Despite its damaging effects to biological molecules, UV light has been shown to be crucial to increase efficiency in prebiotic chemistry. Macromolecular building blocks of life likely require certain levels of surface UV radiation to form (Ranjan & Sasselov 2016; Rimmer et al. 2018). Therefore, whether high UV levels on the surface of planets in the HZ of M stars are a concern or a prerequisite for surface habitability and life for planets orbiting M stars is an open question.

Furthermore, models have shown while UV surface levels for both Proxima b and Trappist-1e are higher than modern Earth, they are lower than the levels early Earth received, even for eroded atmospheres for active input star models (O'Malley-James & Kaltenegger 2019a).

## 4.2 Different planetary surfaces influence the overall reflected planetary flux

For all planetary models shown, we assumed an Earth-like surface (following Kaltenegger, Traub & Jucks 2007). However, exoplanets could have a wide range of possible surfaces, which influences the amount of reflected flux as well as the planet to star contrast ratio. While we do not know which surfaces exist on other Earth-like planets, we explore the effect of different planetary surfaces here for Proxima b.

Fig. 6 shows the reflected spectra of Proxima b for an active stellar phase for 5 different surfaces, representative for major surfaces on Earth assuming (i) the planet is covered by 70% ocean and 30% land, and has a 50% cloud coverage like Earth (Fig. 5 top), and (ii) an idealized cloud-free planet fully covered by one single surface, which shows the maximum impact surfaces can have on a planet's reflected flux (Fig. 6 middle). The individual albedo profiles of each surface are also shown (Fig. 6 bottom).

We also show the effect of clouds on the spectra and contrast ratio as dashed line in Fig. 6. The clouds are assumed to be at 1km, 6km, and 12km, like on Earth (following Kaltenegger et al. 2007). Note that the cloud feedback for Earth-like planets orbiting M stars is still strongly discussed, therefore we have not modified the cloud component in our models (for details see e.g. review Kaltenegger 2017).

We explore the effect of the changing surface on the reflected light to show its effect for different surfaces from an Earth-analog ocean coverage to an idealized one surface no cloud planetary model, which shows the maximum difference due to surface changes. Note that we did not model the planet's climate using the different surfaces, thus the small effect on the surface temperature of the planet's due to different surface reflectivity is not included here.

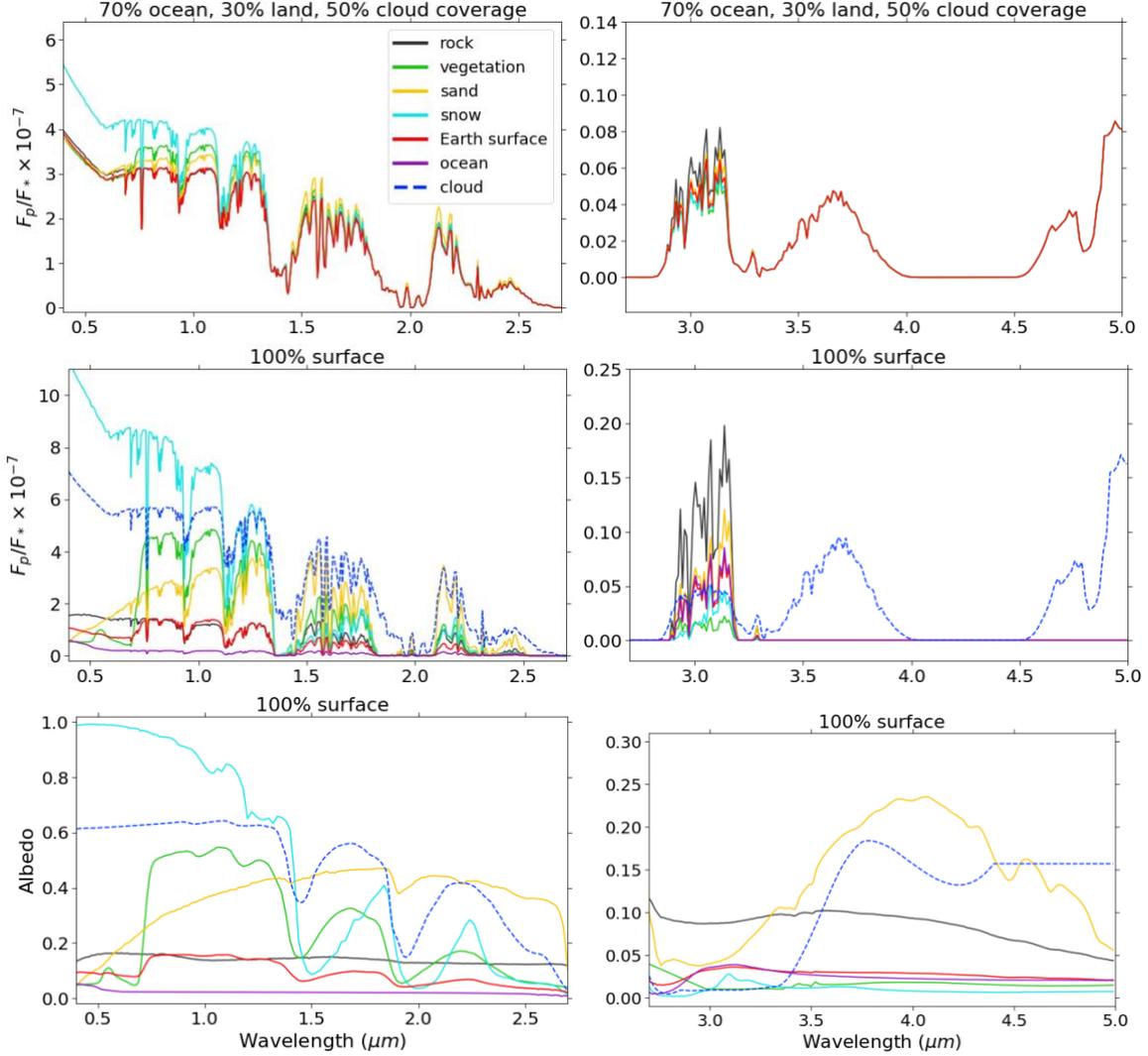

**Fig. 6:** Reflection spectra of Proxima b for 5 different planetary surfaces assuming (top) 70% ocean, 30% land, and 50% cloud coverage, and (middle) an idealized cloud-free single surface planet. The individual albedo profile for each surface component (bottom). Note that spectra from 0.4 to 2.7 μm (left) and spectra from 2.7 to 5 μm (right) are shown in different vertical scales for clarity. All spectra are smeared to a resolving power of 300.

## 5. Conclusions

Here we present high-resolution reflection spectra in the visible to near-infrared (0.4 to 5.0 μm) wavelength range for Proxima b and Trappist-1e, based on different atmospheric models from Earth-like, eroded, to anoxic atmospheres analog to a Young Earth and explore the influence of different surfaces on the reflected spectra.

We find that for both Proxima b and Trappist-1e, spectral absorption features of $H_2O$, $CO_2$, and $CH_4$ are shown for all models with a resolution of 300 in the planet to star contrast ratio as well as the planet's reflected flux. $O_2$ features can be identified for the oxic atmosphere models in the planet to star contrast ratio for both planets. However, to identify $O_2$ features in the planet's reflection spectra, higher resolution is needed (example in Fig. 3 is shown for a resolution of 100,000).

Because of Proxima b's large apparent angular separation, it can be resolved by planned ground-based telescopes like the ELT, increasing the achievable contrast ratio by about $10^3$ compared to unresolved planets. While Trappist-1e cannot be resolved with the ELT, chemical signatures in the atmosphere of unresolved exoplanets have already been observed using high-resolution spectra at the Very Large Telescope, making both planets intriguing targets for future atmospheric characterization in reflected light.

With high-resolution spectrographs such as the HIRES built for the ELT, which has a proposed resolution of 100,000 covering optical to near-infrared wavelengths, the detection of the biosignature combination of $CH_4$ with $O_2$ for oxic atmospheres for both planets is possible in the near future, should it exist on these nearby worlds.

## 6. Acknowledgements

We Thank Jack O'Malley-James for sharing his models and for constructive feedback and insightful discussions. This work was supported by the Carl Sagan Institute and the Breakthrough Foundation.

## References

Airapetian V. S., Glocer A., Khazanov G. V., Loyd R. O. P., France K., Sojka J., Danchi W. C., Liemohn M. W., 2017, ApJ, 836, L3
Anglada-Escudé, G., Amado, P. J., Barnes, J., et al. 2016, Nature, 536, 437
Barnes, R., Deitrick, R., Luger, R., et al. 2016, arXiv:160806919 [astro-ph], http://arxiv.org/abs/1608.06919
Barstow, J. K., & Irwin, P. G. J. 2016, MNRAS, 461, L92
Batalha, N. E., Lewis, N. K., Line, M. R., Valenti, J., & Stevenson, K. 2018, ApJ, 856, L34
Batalha, N. M., Rowe, J. F., Bryson, S. T., et al. 2013, ApJS, 204, 24
Battersby, C., Armus, L., Bergin, E., et al. 2018, Nature Astronomy, 2, 596
Bixel, A., & Apai, D. 2017, ApJ, 836, L31
Borucki, W. J., Agol, E., Fressin, F., et al. 2013, Science, 340, 587
Borucki, W. J., Koch, D. G., Basri, G., et al. 2011, ApJ, 736, 19
Brogi, M., de Kok, R. J., Birkby, J. L., Schwarz, H., & Snellen, I. A. G. 2014, A&A, 565, A124
Cockell C. S., 1998, Theor. Biol., 193, 717
Cockell C. S., Catling D. C., Davis W. L., Snook K., Kepner R. L., Lee P., McKay C. P., 2000, Icarus, 146, 343
Cockell C. S., Kaltenegger L., Raven J. A., 2009, Astrobiol., 9, 623
Dong, C., Lingam, M., Ma, Y., & Cohen, O. 2017, ApJ, 837, L26
Dressing, C. D., & Charbonneau, D. 2015, ApJ, 807, 45
France K. et al., 2016, ApJ, 820, 89
France, K., Froning, C. S., Linsky, J. L., et al. 2013, ApJ, 763, 149
Garcia-Sage K., Glocer A., Drake J. J., Gronoff G., Cohen O., 2017, ApJ, 844, L13
Garraffo C., Drake J. J., Cohen O., 2016, ApJ, 833, L4
Gillon, M., Jehin, E., Lederer, S. M., et al. 2016, Nature, 533, 221
Gillon, M., Triaud, A. H. M. J., Demory, B.-O., et al. 2017, Nature, 542, 456
Goldblatt, C. 2016, arXiv:160807263 [astro-ph], http://arxiv.org/abs/1608.07263
Günther, M. N., Zhan, Z., Seager, S., et al. 2019, arXiv:190100443 [astro-ph],
Heath, M. J., Doyle, L. R., & Joshi, M. M., 20
Hedelt, P., von Paris, P., Godolt, M., et al. 2013, A&A, 553, A9
http://arxiv.org/abs/1901.00443
Husser, T.-O., von Berg, S. W.-, Dreizler, S., et al. 2013, A&A, 553, A6
Jucks, K. W., Johnson, D. G., Chance, K. V., et al. 1998, Geophys Res Lett, 25, 3935
Kaltenegger, L. 2017, Annu Rev Astron Astrophys, 55, 433
Kaltenegger, L., & Sasselov, D. 2010, ApJ, 708, 1162
Kaltenegger, L., & Traub, W. A. 2009, ApJ, 698, 519
Kaltenegger, L., Sasselov, D., & Rugheimer, S. 2013, ApJ, 775, L47
Kaltenegger, L., Segura, A., & Mohanty, S. 2011, ApJ, 733, 35
Kaltenegger, L., Traub, W. A., & Jucks, K. W. 2007, ApJ, 658, 598
Kane SR, Hill ML, Kasting JF, Kopparapu RK, Quintana E V, et al. 2016. Astrophys. J. 830(1):1
Kerwin, B. A., & Remmele, R. L. 2007, Journal of Pharmaceutical Sciences, 96, 1468
Kopparapu R., Wolf E. T., Arney G., Batalha N. E., Haqq-Misra J., Grimm S. L., Heng K., 2017, ApJ, 845, 5
Kreidberg L., Loeb A., 2016, ApJ, 832, L12
Lammer, H., Lichtenegger, H. I. M., Kulikov, Y. N., et al. 2007, Astrobiol., 7, 185
Lederberg J. 1965. Nature 207(4992):9–13
Lincowski, A. P., Meadows, V. S., Crisp, D., et al. 2018, ApJ, 867, 76
Lingam M., Loeb A., 2017, ApJ, 846, L21
Lippincott ER, Eck RV, Dayhoff MO, Sagan C. 1967. Ap. J. 147:753
Lovelock, J. E. 1965. Nature 207(997):568–70
Lovis, C., Snellen, I., Mouillet, D., et al. 2017, A&A, 599, A16
Loyd, R. O. P., France, K., Youngblood, A., et al. 2018, ApJ, 867, 71
Meadows V. S. et al., 2018, Astrobiol., 18, 133
Mennesson, B., Gaudi, S., Seager, S., et al. 2016, in Society of Photo-Optical Instrumentation Engineers (SPIE) Conference Series, Vol. 9904, Proc. SPIE, 99040L
Neale P. J., Thomas B. C., 2016, Astrobiol., 16, 245
O'Malley-James, J. T., & Kaltenegger, L., 2017, MNRAS, 469, L26
O'Malley-James, J. T., & Kaltenegger, L., 2018, MNRAS, 481, 2487
O'Malley-James, J. T., & Kaltenegger, L., 2019a, MNRAS, 485, 5598
O'Malley-James, J. T., & Kaltenegger, L., 2019b, MNRAS, 488, 4530
Onofri S. et al., 2012, Astrobiol., 12, 508
Quintana, E. V., Barclay, T., Raymond, S. N., et al. 2014, Science, 344, 277
Ranjan, S., & Sasselov, D. D. 2016, Astrobiol., 16, 68
Ribas I. et al., 2016, A&A, 596, A111
Ribas, I., Bolmont, E., Selsis, F., et al. 2016, A&A, 596, A111

Rimmer, P. B., Xu, J., Thompson, S. J., et al. 2018, Sci Adv, 4, eaar3302
Rodler, F., & López-Morales, M. 2014, ApJ, 781, 54
Rugheimer S., Segura A., Kaltenegger L., Sasselov D., 2015, ApJ, 806, 137
Rugheimer, S., & Kaltenegger, L. 2018, ApJ, 854, 19
Sancho L. G., de la Torre R., Horneck G., Ascaso C., de los Rios A., Pintado A., Wierzchos J., Schuster M., 2007, Astrobiol., 7, 443
Scalo J. et al., 2007, Astrobiol., 7, 85
See, V., Jardine, M., Vidotto, A. A., et al. 2014, A&A, 570, A99
Segura, A., Walkowicz, L. M., Meadows, V., Kasting, J., & Hawley, S. 2010, Astrobiol., 10, 751
Shields, A. L., Ballard, S., & Johnson, J. A. 2016, Physics Reports, 663, 1
Snellen, I. A. G., de Kok, R. J., le Poole, R., Brogi, M., & Birkby, J. 2013, ApJ, 764, 182
Stevenson, K. B., Lewis, N. K., Bean, J. L., et al. 2016, Publications of the Astronomical Society of the Pacific, 128, 094401
Tarter, J. C., Backus, P. R., Mancinelli, R. L., et al. 2007, Astrobiol., 7, 30
The LUVOIR Team. 2018, arXiv e-prints, arXiv:1809.09668
Tilley, M. A., Segura, A., Meadows, V., Hawley, S., & Davenport, J. 2019, Astrobiol., 19, 64
Traub, W. A., & Stier, M. T. 1976, Appl Opt, 15, 364
Turbet M., Leconte J., Selsis F., Bolmont E., Forget F., Ribas I., Raymond, S. N., Anglada-Escude G., 2016, A&A, 596, A112
Turbet, M., Leconte, J., Selsis, F., et al. 2016, A&A, 596, A112
Udry, S., Bonfils, X., Delfosse, X., et al. 2007, A&A, 469, L43
Vida, K., Kővári, Z., Pál, A., Oláh, K., & Kriskovics, L. 2017, ApJ, 841, 124
Vida, K., Oláh, K., Kővári, Z., et al. 2019, arXiv:190712580 [astro-ph], http://arxiv.org/abs/1907.12580
Vidotto A. A., Jardine M., Morin J., Donati J.-F., Lang P., Russell A. J. B., 2013, A&A, 557, A67
West A. A. et al., 2004, ApJ, 128, 426
West A. A. et al., 2011, Astronom. J., 141, 97
Wolf E., ApJL, 2017, 839, 1
Yamashiki, Y. A., Maehara, H., Airapetian, V., et al. 2019, ApJ, 881, 114
Youngblood A. et al., 2016, ApJ, 824, 101

**Table 1: Stellar and planetary parameters for Proxima b and Trappist-1e and their host stars.** Unless otherwise noted, data are from Gillon et al. 2016, 2017 (for Trappist-1e), and Anglada-Escudé et al. 2016 (for Proxima b).

| STAR | Proxima Centauri | Trappist-1 |
|---|---|---|
| Spectral type | M5.5V | M8.0V |
| $T_{eff}$ (K) | 3050 | 2559 |
| Distance from Earth (pc) | 1.295 | 12.1 |
| Mass ($M_\odot$) | 0.120 | 0.080 |
| Radius ($R_\odot$) | 0.141 | 0.117 |
| Habitable zone range (AU) | 0.024-0.049 | 0.042-0.081 |

| PLANET | Proxima b | Trappist-1e |
|---|---|---|
| Period (d) | 11.186 | 6.0996 |
| Orbital semi-major axis ($10^{-3}$ AU) | 48.5 | 28.17 |
| Minimum Mass ($M_\oplus$) | 1.27 | 0.62 |
| Radius ($R_\oplus$) | 1.07[1] | 0.918 |
| Irradiance compared to Earth | 65% | 66.2% |
| Equilibrium temperature (K) | 234.0 | 251.3 |
| Angular separation[2] (milliarcsec) | 37 | 2.4 |

[1] no data from observation to date, this is a probabilistic estimate assuming rocky composition (Bixel & Apai 2017)
[2] data from O'Malley-James & Kaltenegger 2019a

**Table 2: Input parameters and spectral models used for modeling the climate, photochemistry, and spectra of Proxima b and Trappist-1e.** All data are for the active stellar phase.

| Planet | Proxima b | | | | Trappist-1e | | | |
|---|---|---|---|---|---|---|---|---|
| Model type | Earth-like | Eroded | | Anoxic | Earth-like | Eroded | | Anoxic |
| $P_{surf}$ (bar) | 1 | 0.5 | 0.1 | 1 | 1 | 0.5 | 0.1 | 1 |
| $T_{surf}$ (K) | 278.19 | 268.43 | 252.37 | 275.95 | 271.32 | 262.90 | 250.15 | 292.29 |
| $O_2$ | 20% | 20% | 20% | $3.27 \times 10^{-13}$* | 20% | 20% | 20% | $7.18 \times 10^{-13}$* |
| $CO_2$ | $3.65 \times 10^{-2}$ | $3.65 \times 10^{-2}$ | $3.65 \times 10^{-2}$ | $3.00 \times 10^{-3}$ | $3.65 \times 10^{-2}$ | $3.65 \times 10^{-2}$ | $3.65 \times 10^{-2}$ | $3.00 \times 10^{-3}$ |
| $H_2O_{surf}$ | $5.84 \times 10^{-3}$ | $5.34 \times 10^{-3}$ | $6.25 \times 10^{-3}$ | $8.08 \times 10^{-3}$ | $3.37 \times 10^{-3}$ | $3.24 \times 10^{-3}$ | $5.00 \times 10^{-3}$ | $6.38 \times 10^{-3}$ |

*mixing ratio at surface